\documentclass[preprint,showkeys,preprintnumbers,amsmath,amssymb]{revtex4}

\usepackage{graphicx}
\usepackage{dcolumn}
\usepackage{bm}
\usepackage{natbib}
\usepackage{amssymb}
\usepackage{hyperref}
\usepackage{float}

\begin{document} 

\title{Viscosity of a net-baryon fluid near the QCD critical point}

\author{N.~G.~Antoniou}
\email{nantonio@phys.uoa.gr}
\author{F.~K.~Diakonos}
\email{fdiakono@phys.uoa.gr}
\author{A.~S.~Kapoyannis}
\email{akapog@phys.uoa.gr}
\affiliation{Faculty of Physics, University of Athens, GR-15784 Greece}

\date{5 October 2017}

\begin{abstract}
In the dynamics of the QCD critical point, the net-baryon fluid, 
linked to the slow component of the order parameter, relaxes
to a 3d Ising system in equilibrium.
An analytical study of shear and bulk viscosity, with constraints
imposed by the dynamics of the critical net-baryon fluid, the universality property 
and the requirements of a class of strong coupling theories,
is performed in the neighbourhood of the critical point.
It is found that the shear viscosity of the net-baryon fluid is restricted
in the domain $1.6\leq 4\pi\frac{\eta}{s}\leq 3.7$ for $T_c < T \leq 2T_c$ whereas
the bulk viscosity is small, $4\pi\frac{\zeta}{s} < 0.05$ (for $T>1.27 T_c$) but rising
towards the singularity at $T=T_c$.
\end{abstract}


\maketitle

\section{Introduction}

Near the QCD critical point, the transport coefficients of the strongly
interacting fluid, created in high-energy nuclear collisions, are expected
to exhibit anomalous temperature dependence. This behaviour is implied by
the time evolution of the order parameter fluctuations after reaching
equilibrium. At time scales of the order of the relaxation time the order
parameter fluctuations are governed by the fluctuations of the 
baryon-number density $n_b$. The reason is that the latter is the slow
component of the order parameter due to baryon number conservation 
\cite{Minami,Fujii}.
In contrast, all other degrees of freedom, occurring in the energy-momentum
tensor of the strongly interacting fluid, do not contribute to the
singular behaviour of the transport coefficients in the stage of relaxation
to equilibrium.
In particular the chiral condensate, the
other component of the order parameter, is linked to a fast mode
since the $\sigma$-field becomes massive and therefore does not
participate in the dynamics of the QCD critical point at long time-scales
\cite{Minami}.
As a result, one may claim that although in the study of static properties
of the QCD critical point (critical fluctuations, divergence
of baryon-number susceptibility) both components of the
order parameter $(\sigma,n_b)$ are relevant \cite{Stephanov}, in any attempt to
investigate the dynamic properties of the QCD critical point (divergence of
transport coefficients) only the baryon-number density $n_b$ is a relevant order
parameter \cite{Minami}. Experimentally, one may state that in the search 
for the location of the critical point in the phase diagram, the investigation
of critical fluctuations of $\pi^+\pi^-$ pairs may simulate $\sigma$-field
critical fluctuations \cite{Anticic}. However, dynamic properties of the QCD 
critical point cannot be revealed within a study of transport coefficients
(viscosity) of a meson gas produced in high-energy nuclear collisions
\cite{Prakash}.

In what follows, we consider the behaviour of viscosity (shear and bulk)
in the baryon-number fluid, near the QCD critical point in a process of 
relaxation towards a state of equilibrium described by 3d Ising universality
class. In the same universality class belong also conventional systems
($\rm He$, $\rm N_2$, $\rm H_2O$) of liquid-gas transition \cite{Csernai}
and the basic ingredients in this process out of equilibrium are thermal diffusion
and sound waves \cite{Kadanoff}. The Ising model description in equilibrium
is characterized by the critical exponents of the divergent thermodynamic
quantities and of the correlation length, near the critical point \cite{Huang}.

Within this context, we may conjecture that approaching the critical point,
the following thermodynamic quantities prevail in the description of shear ($\eta$) and bulk
($\zeta$) viscosity: $\eta (T, v_s, \xi, \frac{c_P}{c_V}, \ldots)$, 
$\zeta (\rho, v_s, \xi, \frac{c_P}{c_V}, \ldots)$.
In fact, thermal diffusion produces an inverse relaxation time ($\tau^{-1}$) for a disturbance,
proportional to the specific heat coefficient, $\tau^{-1} \sim c_P$ \cite{Kadanoff} whereas the 
correlation length $\xi$ represents the length scale near the critical point.
The velocity of sound $v_s$ is present since we have assumed that sound waves are among
the nonequilibrium modes in this process.
Finally, from the thermodynamic quantities in the
equation of state, shear viscosity captures the dependence on the temperature ($T$) whereas bulk
viscosity must depend on the mass density ($\rho$) of the medium (net-baryon fluid) in the bulk.
In this approximation, no other quantities are considered in the description of
shear and bulk viscosity, during the relaxation stage.
On the basis, now, of dimensional considerations: [viscosity]=[energy density]$\times$[time] one
may obtain the following expressions in terms of singular quantities in the limit $T \rightarrow T_c$, 
$\mu_b=\mu_c$:
\begin{equation} \label{eq:shear-bulk}
\frac{\eta}{s}=\frac{k_BTv^{-1}_s}{\xi^2s}F^{(s)}\left(\frac{c_P}{c_V} \right) 
\;\; ; \;\;
\frac{\zeta}{s}=\frac{\rho v_s \xi}{s}F^{(b)}\left(\frac{c_P}{c_V} \right)
\end{equation}
where we have introduced the entropy density ($s$) forming the dimensionless ratios 
(\ref{eq:shear-bulk}) in the system of units $k_B=c=\hbar=1$. 
The basic thermodynamics of the fluid is formulated in terms
of the relations:
\begin{equation} \label{eq:basicthermo}
c_P-c_V=Tk_T \left( \frac{\partial P}{\partial T}\right) ^2_V  \;\; , \;\;
\frac{c_P}{c_V}=\frac{k_T}{k_S} \;\; , \;\;
v^2_s=(\rho k_S)^{-1} \;\; , \;\;
s=\frac{\varepsilon+P}{T}-\frac{\mu_b n_b}{T}
\end{equation}
where $k_T$, $k_S$ are the isothermal and adiabatic (isentropic)
compressibility, $\varepsilon$ the energy density, $P$ the pressure
and $\mu_b$ the baryochemical potential.

A minimal requirement of relativistic thermodynamics leads to the 
identification of the mass density $\rho$ of the fluid with the enthalpy
density: $h=\varepsilon + P$ ($\rho=\frac{h}{c^2}$) in eqs.~(\ref{eq:shear-bulk}) 
and (\ref{eq:basicthermo}) as a result of the properties of the energy-momentum
tensor \cite{Landau}. Moreover, in order to fix the amplitudes (scales)
of the singular quantities, consistently with relativity, one may consider
the net baryon-fluid consisting, in the quark phase ($T  \gg T_c$),
of the quark-excess with conserved number density, following 
the equation of state of an ideal, massless, classical system and leading to:
\[
\varepsilon=3P \;\;,\;\;
P=n_b T \;\;,\;\;
h=4n_b T \;\;,\;\;
c_V=3n_b \;\;,\;\;
c_P=4n_b \;\;,
\]
\begin{equation} \label{eq:ideal}
k_T=(n_bT)^{-1} \;\;,\;\;
s=\left(4-\frac{\mu_b}{T} \right) n_b \;\;,\;\;
v_s=\frac{1}{\sqrt{3}}
\end{equation}
In this description the remaining balanced quarks and antiquarks (zero baryon number) form an environment for the net-baryon fluid which affects the dependence of the chemical potential $\mu_b$ on the temperature $T$.

\section{Critical exponents of viscosity}

Introducing now the appropriate critical exponents and the corresponding
amplitudes, we obtain in the limit $T \rightarrow T_c$, $\mu_b=\mu_c$,
the power laws:
\begin{equation} \label{eq:powerlaws}
c_V =A_\pm |t|^{-\alpha} ,
k_T=\Gamma_\pm |t|^{-\gamma} ,
\xi=\xi_\pm |t|^{-\nu} \;
\left( t \equiv \frac{T-T_c}{T_c}  \right)
\end{equation}
where the indices ($\pm$) in the amplitudes correspond to the limits
$t \rightarrow 0^+$ and $t \rightarrow 0^-$ respectively \cite{Huang}.

In eqs.~(\ref{eq:powerlaws}) not only the critical exponents are universal
but also the ratios of the amplitudes $\frac{A_+}{A_-}$, 
$\frac{\Gamma_+}{\Gamma_-}$, $\frac{\xi_+}{\xi_-}$, corresponding
to the phases $T>T_c$ (net-baryon, quark-matter fluid) and $T<T_c$
(net-baryon, baryonic fluid), are fixed within the universality class of 
the critical point \cite{Huang}. Moreover, following our discussion above,
the amplitudes ($A_+$, $\Gamma_+$) in the quark-matter phase, representing
the scales of the thermodynamic quantities ($c_V$, $k_T$) near the critical temperature, can be fixed with the 
help of eqs.~(\ref{eq:ideal}) assuming a continuous transition to the ideal behaviour. 
To ensure the continuity of the sound velocity (see eqs.~(\ref{eq:shearbulkampl}) below), reaching the 
constant value $\frac{1}{\sqrt{3}}$ in the ideal regime, the matching of $c_V$, $k_T$ 
has to be taken at $t=1$ ($T=2T_c$) leading to the relations:
\begin{equation} \label{eq:amplplus}
A_+=3n_c \;\;,\;\;
\Gamma_+ = (2n_c T_c)^{-1}
\end{equation}
where $n_c \equiv n_b(T_c) = n_b(2T_c)$ denotes the critical baryon-number density. Here we have made use 
of the fact that the order parameter $n_b(T)-n_c$ vanishes in the symmetric phase $T>T_c$ \cite{Antoniou}.
We are aware of the fact that the extrapolation of the power laws (\ref{eq:powerlaws}),
beyond the critical region, is a crude approximation.
Our conjecture, however, is that the combination of the singular thermodynamic
quantities to form the expressions (\ref{eq:shear-bulk}) of shear and bulk viscosity
may lead to a solution, valid also in a distance from
$T_c$, developing a noncritical behaviour there.
Obviously, this conjecture can only be verified a posteriori, at the end of our treatment.

In this framework one may proceed to a semi-quantitative treatment of shear
and bulk viscosity, near the QCD critical point, on the basis of 
eqs.~(\ref{eq:shear-bulk}, \ref{eq:basicthermo}, \ref{eq:powerlaws}).
For the functions $F^{(i)}\left(\frac{c_P}{c_V} \right)$, $i:(s,b)$ in eqs.~(\ref{eq:shear-bulk}) 
we adopt a simple model inspired by a perturbative
treatment of conventional fluids, in the vicinity of liquid-gas
critical point, sharing the same universality class (3d Ising) with QCD \cite{Kadanoff}: 
$F^{(i)}\left(\frac{c_P}{c_V} \right)=f^{(i)}\frac{c_P}{c_V}$
where the dimensionless constants $f^{(i)}$ are not universal, they depend 
on the nature and the length scale of the medium at microscopic level.
With this choice, equations (\ref{eq:shear-bulk}, \ref{eq:basicthermo})
give:
\[
v^2_s=\frac{k_T^{-1}}{h} 
\left[ 1+Tk_T \left( \frac{\partial P}{\partial T} \right)^2_V c_V^{-1} \right]\;\;,
\]
\[
\frac{\eta}{s}=f^{(s)} \frac{T^{3/2}h^{1/2}}{s}
\left( \frac{\partial P}{\partial T}\right)_V
k_T \xi^{-2} c_V^{-1/2}  
\left[1+T^{-1} \left( \frac{\partial P}{\partial T}\right)^{-2}_V 
\frac{c_V}{k_T} \right]^{1/2}\;\;,
\]
\begin{equation} \label{eq:vshearbulk}
\frac{\zeta}{s}=f^{(b)} \frac{T^{3/2}h^{1/2}}{s}
\left( \frac{\partial P}{\partial T}\right)^3_V
k_T \xi c_V^{-3/2}
\left[1+T^{-1} \left( \frac{\partial P}{\partial T}\right)^{-2}_V 
\frac{c_V}{k_T} \right]^{3/2}
\end{equation}
Incorporating the power laws (\ref{eq:powerlaws}) for $c_V$, $k_T$ and $\xi$
in eqs.~(\ref{eq:vshearbulk}) we obtain the singular forms in the limit
$T \rightarrow T_c$:
\[
v^2_s=\frac{|t|^{\alpha}}{h} 
\left[ \Gamma^{-1}_{\pm} |t|^{\gamma-\alpha} +
 A^{-1}_{\pm} T \left( \frac{\partial P}{\partial T} \right)^2_V \right]\;\;,
\]
\[ 
\left( \frac{\eta}{s} \right)_{\pm} =
f^{(s)} \frac{T_c^{3/2}h_c^{1/2}\lambda_c}{s_c}
\left( \Gamma_{\pm} \xi^{-2}_{\pm} A^{-1/2}_{\pm} \right)
\left(1+T^{-1}_c \lambda_c^{-2} A_{\pm} \Gamma_{\pm}^{-1} |t|^{\gamma-\alpha}
\right)^{1/2}
|t|^{-\gamma+2\nu+\frac{\alpha}{2}}\;\;,
\]
\begin{equation} \label{eq:shearbulkampl}
\left( \frac{\zeta}{s} \right)_{\pm} =
f^{(b)} \frac{T_c^{3/2}h_c^{1/2}\lambda_c^3}{s_c}
\left( \Gamma_{\pm} \xi_{\pm} A^{-3/2}_{\pm} \right)
\left(1+T^{-1}_c \lambda_c^{-2} A_{\pm} \Gamma_{\pm}^{-1} |t|^{\gamma-\alpha}
\right)^{3/2}
|t|^{-\gamma-\nu+\frac{3\alpha}{2}}
\end{equation}
where $\lambda_c \equiv \left( \frac{\partial P}{\partial T} \right)_V$ at
$T=T_c$. In fact eqs.~(\ref{eq:shearbulkampl}) contain also a non-critical contribution
increasing for $T>T_c$. This leads to a solution interpolating smoothly between the critical
and the asymptotic ideal gas behaviour, taking also the constraint of
baryon-number conservation into account.
The critical exponents ($\alpha,\gamma,\nu$) are not
independent since they are constrained by the Josephson scaling law $\nu d = 2-\alpha$. Therefore,
the indices of the power laws in eq.~(\ref{eq:shearbulkampl}) are given in terms
of two independent critical exponents. In particular, the leading power laws
of shear and bulk viscosity are:
\begin{equation} \label{eq:shearbulklead}
\eta \sim |t|^{1-\gamma+\frac{\nu}{2}} \;\;,\;\;
\zeta \sim |t|^{3-\gamma-\frac{11}{2}\nu}
\end{equation}
where the exponents ($\gamma,\nu$) are expected to be compatible with Ising-like
universality class in 3d. In fact the behaviour (\ref{eq:shearbulklead}) is
universal, it is valid near the liquid-gas critical point of conventional matter
\cite{Kadanoff} and also near the quark-hadron critical point of QCD matter.
On the other hand, the behaviour (\ref{eq:shearbulklead}) must be in accordance with the
dynamical aspects of the QCD critical point which suggest, according to a compilation of predictions
\cite{Minami,Monnai,Son,Moore,Onuki,Hohenberg}, the singular behaviour
$\eta\sim \xi^{0.05}$, $\zeta\sim \xi^{2.8}$.
Comparing these power laws with the behaviour (\ref{eq:shearbulklead}) we find ($\xi \sim |t|^{-\nu}$):
$\gamma \simeq 1.34$, $\nu \simeq 0.61$, a solution compatible with Ising-like
universality class.

Finally, in order to verify explicitly, the compatibility of the scaling relations
(\ref{eq:shearbulklead})
with the dynamics of the critical point, we consider the treatment in ref.~\cite{Onuki}
in which the renormalized transport coefficients, computed in the $\epsilon$-expansion
behave as follows: $\eta_R \sim \xi ^{\epsilon /19}$ and $\zeta_R (0) \sim \xi^{z-\alpha/\nu}$
in the slow mode $\omega=0$. The dynamic critical exponent $z$ is given by the expansion
$z=4-\frac{18}{19}\epsilon+\ldots$ and for a 3d fluid and the Ising exponents $\nu \simeq 0.61$,
$\alpha = 2-\nu d$, one finds, to first order in $\epsilon=4-d$ : $\eta_R \sim \xi^{0.053}$,
$\zeta_R (0) \sim\xi^{2.77}$, in a very good agreement with eqs.~(\ref{eq:shearbulklead}).

\section{Singular solutions}

The characteristic properties
of viscosity near the QCD critical point, described by the solution
(\ref{eq:shearbulkampl}), depend on a number of nonuniversal amplitudes which are
fixed by the following constraints:

(a) the assumption that in the quark-matter phase {($T \gtrsim T_c$)} the amplitudes
$\Gamma_+$, $A_+$ are given by eqs.~(\ref{eq:amplplus}) which are compatible with the 
equation of state of a non-interacting, massless, classical system with constant baryon-number density
$n_b(T)=n_c$ for $T \gg T_c$ and

(b) the universality constraint imposed on the ratios of the Ising amplitudes:
$\frac{A_+}{A_-}=0.5-0.6$, $\frac{\xi_+}{\xi_-}=2$, $\frac{\Gamma_+}{\Gamma_-}=4.5-5.0$
\cite{Privmam}.

In fact, the solution (\ref{eq:shearbulkampl}) can be written in a simplified form:
\[
\left( \frac{\eta}{s} \right)_{\pm} =
f^{(s)} M_{\pm} 
\left(1+\Lambda_{\pm} |t|^{\gamma+3\nu-2} \right)^{1/2}
|t|^{1-\gamma+\frac{\nu}{2}}
\]
\begin{equation} \label{eq:shearbulksimpl}
\left( \frac{\zeta}{s} \right)_{\pm} =
f^{(b)} N_{\pm} 
\left(1+\Lambda_{\pm} |t|^{\gamma+3\nu-2} \right)^{3/2}
|t|^{3-\gamma-\frac{11}{2}\nu}
\end{equation}
where:
\[
M_{\pm}\equiv
\frac{T_c^{3/2}h_c^{1/2}\lambda_c}{s_c}
\Gamma_{\pm} \xi^{-2}_{\pm} A^{-1/2}_{\pm} \;\;,\;\;
\]
\[
N_{\pm}\equiv
\frac{T_c^{3/2}h_c^{1/2}\lambda_c^3}{s_c}
\Gamma_{\pm} \xi_{\pm} A^{-3/2}_{\pm} \;\; {\rm and}
\]
\begin{equation} \label{eq:amplitudes}
\Lambda_{\pm}\equiv
T_c^{-1}\lambda_c^{-2}A_{\pm}\Gamma_{\pm}^{-1} \;\;,
\end{equation}
with $h_c = 4 n_c T_c$, $\lambda_c=n_c$, 
$s_c=\left(4-\frac{\mu_c}{T_c} \right)n_c$ assuming the (approximate) validity of eqs.~(\ref{eq:ideal})
in the temperature range $(T_c,2T_c)$ for all thermodynamic quantities which do not possess divergent 
singularities for $T \rightarrow T_c$. 
The constraints (a) and (b) lead to the following solution for the
dimensionless amplitudes (\ref{eq:amplitudes}):
\[
M_+ = \frac{1}{\sqrt{3}} \frac{\xi_+ ^{-2} T_c}{s_c} \;\;,\;\;
N_+ = \frac{1}{3\sqrt{3}} \frac{n_c \xi_+ T_c}{s_c} \;\;,\;\;
\Lambda_+ =6
\]
\begin{equation} \label{eq:amplitudevalues}
M_- = \frac{4}{7\sqrt{3}} \frac{\xi_+ ^{-2} T_c}{s_c} \;\;,\;\;
N_- = \frac{1}{84\sqrt{3}} \frac{n_c \xi_+ T_c}{s_c} \;\;,\;\;
\Lambda_- =60
\end{equation}
where $\xi_+ \simeq$ 1 fm a typical scale of the correlation length and a set of critical
values ($T_c$, $\mu_c$, $n_c$) can be taken from reference \cite{Antoniou} where a study of 
baryon-number susceptibility near the critical point is performed and also from NA49 measurements
in a search for critical fluctuations \cite{Anticic,NA49}: $T_c \simeq$ 160 MeV, $\mu_c \simeq$
220 MeV, $n_c \simeq$ 0.13 fm$^{\rm -3}$. The location of the critical point
($T_c,\mu_c$) in the QCD phase diagram is still an open problem and the above
values are only indicative. However a change of these quantities affects only the
actual values of the constants $f^{(i)}$ but not the solution (\ref{eq:shearbulksimpl}). In fact,
if we choose the critical values suggested by
Lattice QCD: $T_c \simeq$ 160 MeV, $\mu_c \simeq$ 400 MeV \cite{Fodor,Ejiri}, the constants
$f^{(i)}$ increase by a factor 1.7 but the overall prefactors 
$f^{(s)} M_{\pm}$, $f^{(b)} N_{\pm}$ in eq.~(\ref{eq:shearbulksimpl}) remain
unchanged, fixed by the constraint of Kovtun-Son-Starinets (KSS) bound, described in the discussion below.

To complete our treatment and determine the remaining constants
$f^{(s)}$, $f^{(b)}$ in eqs.~(\ref{eq:shearbulksimpl}), we employ as a
final guiding principle, the KSS bound \cite{Kovtun} which is assumed
to be reached by the minimum of the ratio $\frac{\eta}{s}$, located very
close to the critical temperature, in the hadronic phase ($t \simeq -2.9 \cdot 10^{-3}$)
according to eqs.~(\ref{eq:shearbulksimpl}).
This constraint has its origin in a class of strong coupling field
theories (anti-de Sitter/conformal field theory (AdS/CFT) limit)
and it is widely accepted that the formation of
quark matter in high-energy nuclear collisions creates an ideal
environment in order to test its validity \cite{Csernai}. Also, in the
same framework, a constraint on the bulk viscosity can be
obtained if we use the parametrization, 
$\frac{\zeta}{s}=\frac{1}{8\pi}\left(\frac{1}{3}-v_s^2\right)$, introduced in reference 
\cite{Noronha-Hostler} and take, for our purpose, the average in the domain
$0.5 \leq t \leq 1$. From eqs.~(\ref{eq:shearbulkampl}) we have for
$T_c \leq T \leq 2T_c$:
\begin{equation} \label{eq:vsresult}
v_s^2=\frac{t^{2-3\nu}}{4}
\left( \frac{2t^{\gamma+3\nu-2}}{1+t}+\frac{1}{3} \right) \;\;;\;\;
\left< v_s^2 \right> \simeq 0.27
\end{equation}
Thus, we obtain the final constraints:
\begin{equation} \label{eq:shearbulkresult}
\left( \frac{\eta}{s} \right)_{\rm min} = \frac{1}{4\pi}\;\;(t\simeq 0^-) \;\;;\;\;
\left< \left( \frac{\zeta}{s} \right)_+ \right> \simeq \frac{0.030}{4\pi}
\end{equation}
which lead to the estimate: $f^{(s)}\simeq$ 8.2$\times$10$^{-2}$ and
$f^{(b)}\simeq$ 2.0$\times$10$^{-3}$.

\section{Discussion and conclusions}

\begin{figure}[H]
\centering
\vspace{-2.5cm}
\hspace{-0.7cm}
\includegraphics[scale=0.8,angle=-0]{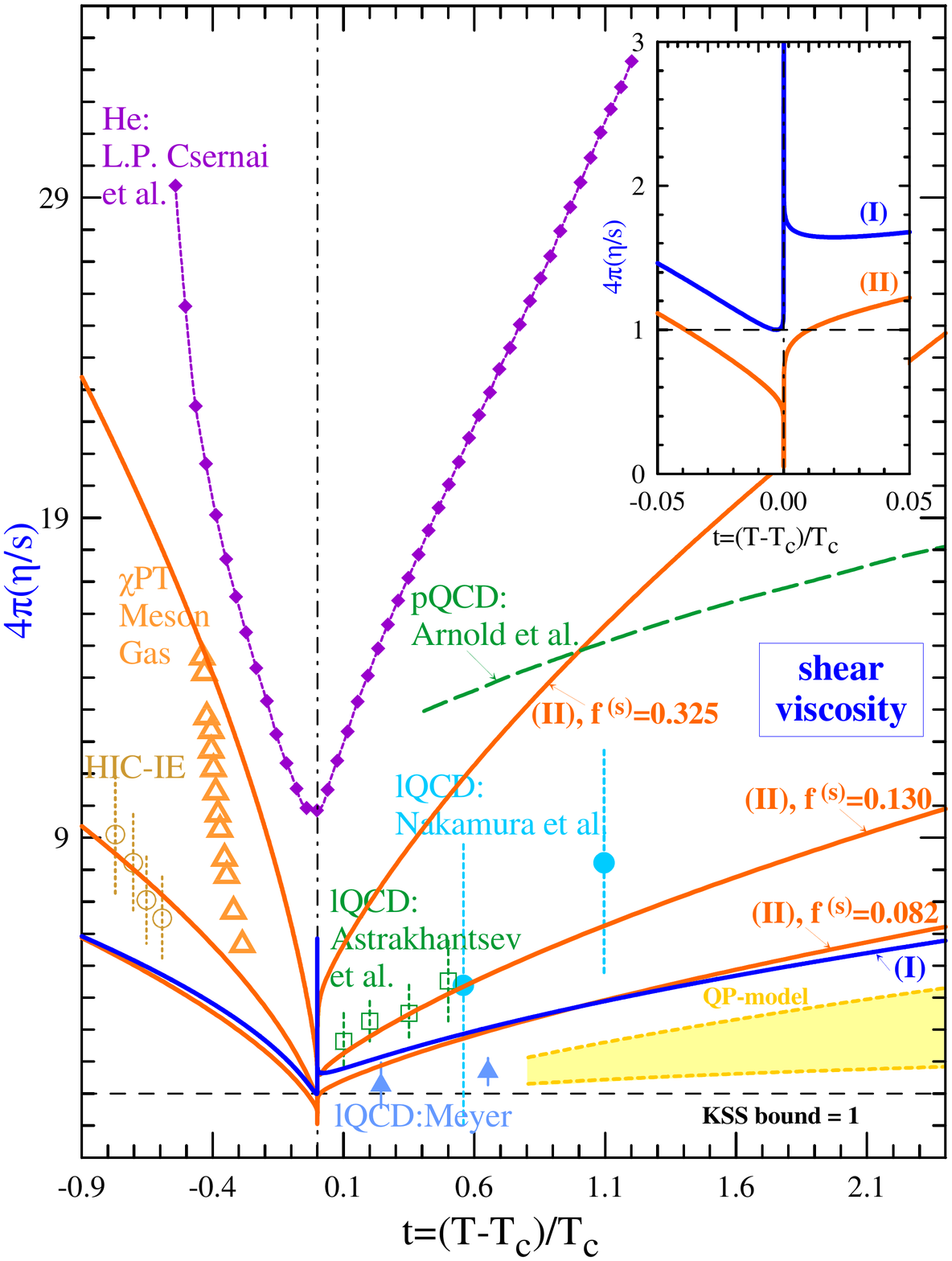}
\vspace{-2cm}
\caption{\label{fig:shear} Our solution of type I (continuous dark line) for the shear viscosity
compared with the findings of \cite{Prakash} (empty triangles), 
\cite{Danielewicz} (empty circles), 
\cite{Nakamura} (solid circles and solid triangles),
\cite{Astrakhantsev} (empty rectangles),
\cite{Arnold} (dashed line), a quasi-particle model \cite{Plumari} (band) and
\cite{Csernai} (dotted line with solid rectangles). In the inset graph we focus on
the shape of our solution in the vicinity of the critical temperature.
Also solutions of type II (continuous light lines) are shown.}
\end{figure}

The solution for shear viscosity is unstable at $T=T_c$, under small changes of
the critical exponents ($\nu,\gamma$). In fact if we consider the values of 3d
Ising exponents, given by recent theoretical studies \cite{Gliozzia} $\nu \simeq 0.63$,
$\gamma \simeq 1.24$, the weakly divergent singularity at $T=T_c$ discussed above
(solution I), becomes a cusp singularity (solution II). 
This solution vanishes at $T=T_c$ but, for the same value of $f^{(s)}$, 
it approaches rapidly the solution I as we
depart from the critical temperature ($|t| \gg 0.04$), leaving, as a single imprint,
a cusp at $t=0$. 
The solution II violates the KSS bound and therefore the dimensionless constant
$f^{(s)}$ remains a free parameter. In Fig.~\ref{fig:shear} we show, for illustration,
three different solutions of the type II, sharing the same cusp singularity and corresponding
to the values $f^{(s)}=$ 0.082, 0.130 and 0.325. In fact, solutions of the type I and II are presented in
Fig.~\ref{fig:shear} and compared with other findings, not related to critical behaviour.
In the hadronic phase
($T<T_c$) we found $1 \leq 4\pi\frac{\eta}{s} \leq 4.3$ for $\frac{T_c}{2} \leq T < T_c$ (in solution I),
deviating from the behaviour of chiral matter (meson
gas in chiral perturbation theory) \cite{Prakash} and the behaviour of
$\frac{\eta}{s}$ extracted from heavy-ion collisions at intermediate energies
(HIC-IE) \cite{Danielewicz}.
For $T>T_c$ (quark matter) we found $1.6 \leq 4\pi\frac{\eta}{s} \leq 3.7$ for 
$T_c < T \leq 2T_c$ (in solution I) and a comparison with recent results of lattice QCD (lQCD) for the
shear viscosity of gluonic matter \cite{Nakamura} is illustrated.
In the same figure, it is of particular interest to compare our solutions with the
estimate of the ratio $\frac{\eta}{s}$ for QCD with dynamical $N_f=3$ quarks
given in reference \cite{Astrakhantsev}. Also, in Fig.~\ref{fig:shear},
predictions of perturbative QCD \cite{Arnold} and of a quasi-particle model \cite{Plumari} (band) are
presented for comparison. Finally, it is of interest to note that the weakness of the
singularity, at $T=T_c$, manifests itself as a two-minima structure, very close to the critical
temperature (Fig.~\ref{fig:shear}). The absolute minimum reaches the KSS bound in the hadronic
and not in the quark-matter phase. This structure cannot be seen in the coarse data of
conventional matter, as shown in Fig.~\ref{fig:shear} in the case of helium \cite{Csernai} and,
certainly, is not expected to be observable in high-energy nuclear collisions either. 
We also observe that in the quark-matter phase ($T>T_c$) the critical behaviour
of shear viscosity is confined in a very narrow region
$\frac{\Delta T}{T_c} \simeq 10^{-2}$ corresponding to the position of the local minimum at 
$t_{min} \simeq 2.0 \cdot 10^{-2}$.
Departing from this region ($t \gg 10^{-2}$) the solution (\ref{eq:shearbulksimpl}) may
still be valid, dominated by the square-root term which leads to a smoothly
increasing function with a noncritical behaviour (Fig.~\ref{fig:shear}).
Moreover, in a distance from the critical point
($t \simeq 1$) the properties of shear viscosity are

\begin{figure}[h]
\centering

\hspace{-0.6cm}
\includegraphics[scale=0.57,angle=-0]{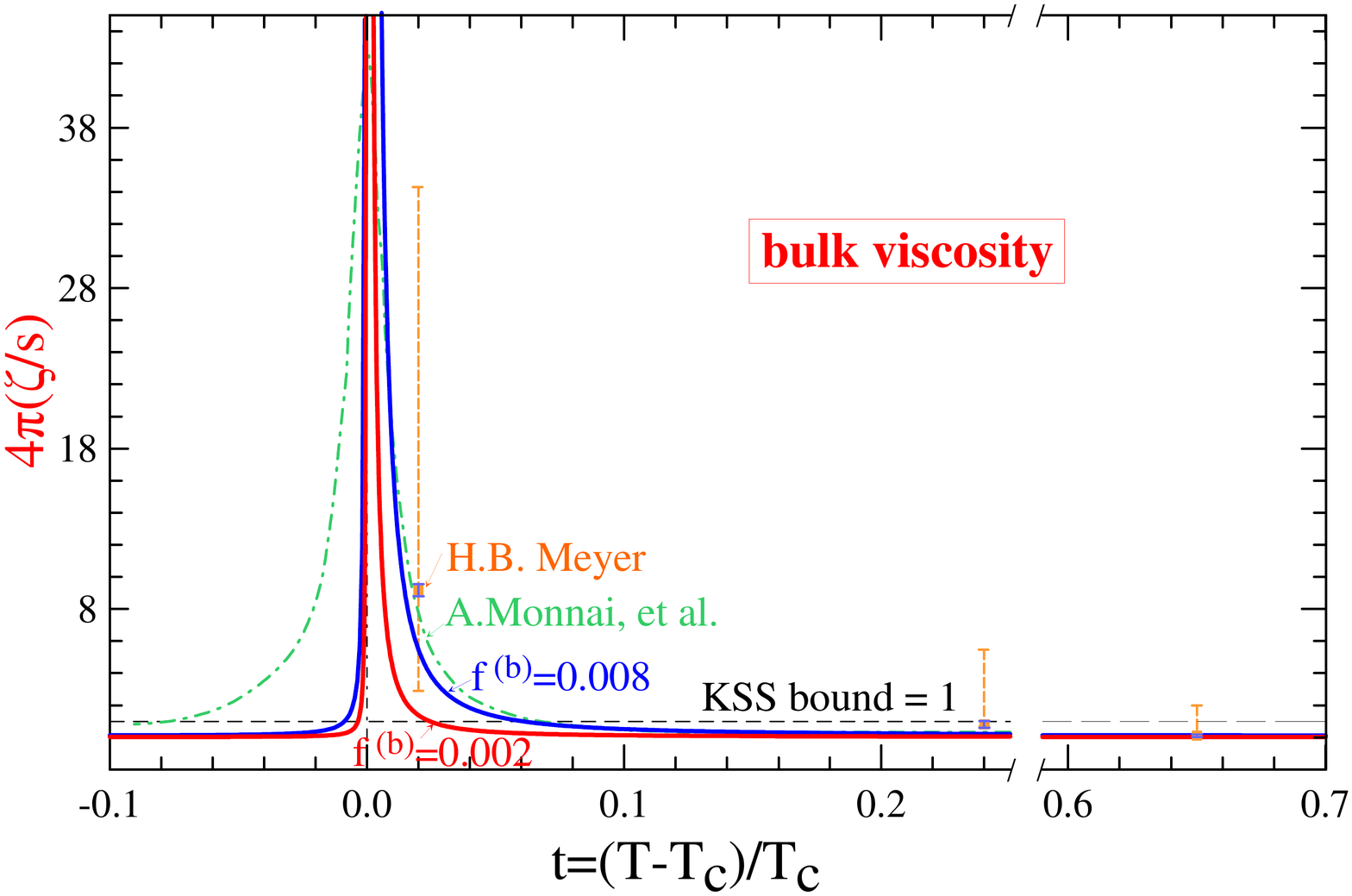}

\vspace{-1cm}
\caption{\label{fig:bulk} Solutions for the bulk viscosity (continuous lines) 
compared with the findings of \cite{Monnai} (dot-dashed line) and 
\cite{Meyer} (solid rectangles) with systematic (large) and statistical
(small) uncertainties.}
\end{figure}

\noindent
expected to deviate
from the requirement of strong coupling regime and come
close to the properties of a non-interacting system with conserved
baryon-number density (\ref{eq:ideal}).
This observation justifies a posteriori the constraint
(\ref{eq:amplplus}) on the amplitudes 
$A_+$, $\Gamma_+$. 
Similar remarks may apply to bulk viscosity which remains practically constant beyond 
its critical region
$\frac{\Delta T}{T_c} \simeq 10^{-2}$ (Fig.~\ref{fig:bulk}).

In Fig.~\ref{fig:bulk} the bulk viscosity of net-baryon matter, in our solution, is presented
(continuous light line). The behaviour of bulk viscosity remains practically unchanged 
under small changes of the critical exponents ($\nu,\gamma$).
Despite the fact that it develops a strong singularity at the critical
point ($\zeta \sim \xi^{2.8}$) it decreases rapidly and stays to an approximately constant value,
smaller than the KSS bound, for $|t| \gg 0.025$. Our result is compared with the solution (dot-dashed
line) in reference \cite{Monnai} where a dynamical treatment of enhanced bulk viscosity near the
critical point is performed. In the same figure, the results of lattice QCD
for gluonic matter are shown \cite{Meyer} whereas in a similar lQCD treatment
\cite{Danielewicz} the results for the bulk viscosity of gluonic matter are
compatible with zero, for $\frac{3T_c}{2} < T < 2T_c$, and are not shown in
Fig.~\ref{fig:bulk}.

Finally, if we remove the constraint on bulk viscosity, inspired by gauge/gravity
duality, given in eq.~(\ref{eq:shearbulkresult}), the constant $f^{(b)}$ remains
a free parameter. In Fig.~\ref{fig:bulk} a solution with $f^{(b)}=$ 0.008
is also shown (continuous dark line), approaching, in the quark phase, the solution described 
in ref.~\cite{Monnai}.

In summary, an analytical study of shear and bulk viscosity of net-baryon matter
near the QCD critical point is performed. It is based on the assumption that
the net-baryon fluid, associated with the slow order parameter
(baryon-number density $n_b$) of the critical phenomenon, relaxes, in a
process out of equilibrium, to the Ising universality class in equilibrium.
The universal indices (critical exponents, ratios of the amplitudes of critical
singularities) are basic ingredients in this approach, leading to a prediction
of viscosity near the critical point. 
This becomes possible if we employ constraints inspired by gauge/gravity
duality, in the strong-coupling regime, near the critical point (KSS bound for
shear viscosity and a related parametrization for bulk viscosity).
In fact, these constraints provide us with an estimate of the dimensionless
constants $f^{(s)}$, $f^{(b)}$ (eq.~\ref{eq:shearbulkresult}) something which
at present is beyond our capability to calculate in QCD. If we remove these
constraints, these constants remain free parameters.
As a final conclusion, this study suggests that precision measurements
of elliptic flow of net protons at the SPS (NA61 experiment) or at
RHIC in the Beam Energy Scan program \cite{STAR}, are of particular importance
since they are strongly linked to the dynamics of the QCD critical point.

\end{document}